\documentclass[prl,twocolumn,superscriptaddress,showpacs,floats,floatfix]{revtex4}
\usepackage{epsfig,dcolumn,amsmath,latexsym}

\begin{document}

\title{Absence of magnetically-induced fractional quantization in atomic contacts}

\preprint{7}

\author{C.\ Untiedt}
\affiliation{Kamerlingh Onnes Laboratorium, Universiteit Leiden,
Postbus 9504, NL-2300 RA Leiden, The Netherlands}
\affiliation{Departamento de F\'\i sica Aplicada, Universidad de
Alicante, Campus de San Vicente del Raspeig, E-03690 Alicante,
Spain.}

\author{D.\,M.\,T.\ Dekker}
\affiliation{Kamerlingh Onnes Laboratorium, Universiteit Leiden,
Postbus 9504, NL-2300 RA Leiden, The Netherlands}

\author{D.\ Djukic }
\affiliation{Kamerlingh Onnes Laboratorium, Universiteit Leiden,
Postbus 9504, NL-2300 RA Leiden, The Netherlands}

\author{J.\,M.\ van Ruitenbeek}
\affiliation{Kamerlingh Onnes Laboratorium, Universiteit Leiden,
Postbus 9504, NL-2300 RA Leiden, The Netherlands}

\date{\today}

\begin{abstract}
Using the mechanically controlled break junction technique at low
temperatures and under cryogenic vacuum conditions we have studied
atomic contacts of several magnetic (Fe, Co and Ni) and
non-magnetic (Pt) metals, which recently were claimed to show
fractional conductance quantization. In the case of pure metals we
see no quantization of the conductance nor half-quantization, even
when high magnetic fields are applied. On the other hand, features
in the conductance similar to (fractional) quantization are
observed when the contact is exposed to gas molecules.
Furthermore, the absence of fractional quantization when the
contact is bridged by H$_2$ indicates the current is never fully
polarized for the metals studied here. Our results are in
agreement with recent model calculations.
\end{abstract}

\pacs{75.75.+a, 73.63.Rt}

\maketitle

When a metallic wire is stretched its conductance becomes smaller
as a result of the decrease of its cross section. This process
continues until the breaking of the wire, and just before this
event takes place, the contact is formed by just one atom. In this
way atomic-sized contacts between two metallic electrodes can be
formed and studied. The instruments that have made these studies
possible are the mechanically controllable break junctions and the
scanning tunnelling microscope. In both techniques the relative
displacement of two electrodes is controlled with a resolution of
a few picometers by the use of a piezoelectric element which
allows to monitor the formation and breaking of the contact
between the two electrodes.

Properties of such atomic-sized contacts have been extensively
studied during the past decade \cite{agrait03} for many different
metals both magnetic and non-magnetic. The conductance of these
contacts can be described by the Landauer formula:
\begin{equation}
G=G_0 \sum_i T_i
\end{equation}
where the summation is extended to all the available channels for
the electrons traversing the contact, $T_i$ is a number between 0
and 1 for the transmission of the $i^{th}$ channel and
$G_0=2e^2/h$ the quantum of conductance (assuming degeneracy of
spin) in terms of the electron charge $e$ and Planck's constant
$h$.  In the case in which the degeneracy of spin would be removed
the channels would have to be redefined for each spin and each of
these would carry up to $\frac{1}{2}G_0$.

The number of channels available in a one-atom contact is
determined by the valence of the metal \cite{scheer98}, and the
transmission of each channel is fixed by other parameters such as
the number of neighbors or the bond distance
\cite{scheer97,cuevas98}. For special cases ({\it s}-type metals
like Au or Na), electronic transport through a single atom will be
due to a single channel with a transmission close to unity, but
this will not be true for other metals where all kinds of
combinations of channels with different transmissions will add up
to produce the total conductance of the atom. This is the case for
transition metals with partial occupation of the d-orbitals and
therefore they are not expected to have a one-atom conductance of
$\frac{1}{2} G_0$ or $1\,G_0$ for magnetic or non-magnetic metals,
respectively \cite{delin03,bagrets03}. This will make it difficult
to establish just from the conductance whether the spin-degeneracy
of the conductance channels has been lifted or not. However,
several claims have appeared of the observation of half-integer
conductance quantization for both magnetic and non-magnetic metals
\cite{elhoussine02,shimizu02,gillingham02,gillingham03,gillingham03a,rodrigues03}.
The claims are based on the observation of peaks in conductance
histograms at half integers of the quantum of conductance in
experiments made at room temperature. These observations cannot be
understood from the present knowledge of transport properties of
atomic-sized contacts. Especially the claim of the observation of
values of conductance at half integers of the quantum in the case
of Pt and the interpretation of this phenomenon as the result of
the lifting of the spin degeneracy \cite{rodrigues03} seem to
contradict previous theoretical \cite{delin03} and experimental
\cite{sirvent96a,smit01,nielsen03} work.

To further investigate this problem we have studied the
conductance of several magnetic metals, namely Fe, Co and Ni, and
a non-magnetic one, Pt. We have used the mechanically controllable
break junction technique that uses a notched wire of the metal
under study, which is glued to both sides of the notch on top of a
bending beam. By bending of the beam the wire is broken at the
notch and with the use of a piezo-element the relative
displacement of the two resulting electrodes can be controlled
with a precision of a few picometers. The first breaking of the
wire is done only at 4.2K under cryogenic vacuum in order to
assure that the atomic contact is formed between clean electrode
surfaces. The break junction device remains under these conditions
for the full duration of the experiment. The starting purity of
our samples was 99.998\,\% for Pt, Fe and Ni and of 99.99\,\% for
Co.

\begin{figure}[!t]
\begin{center}
\epsfig{width=7cm,figure=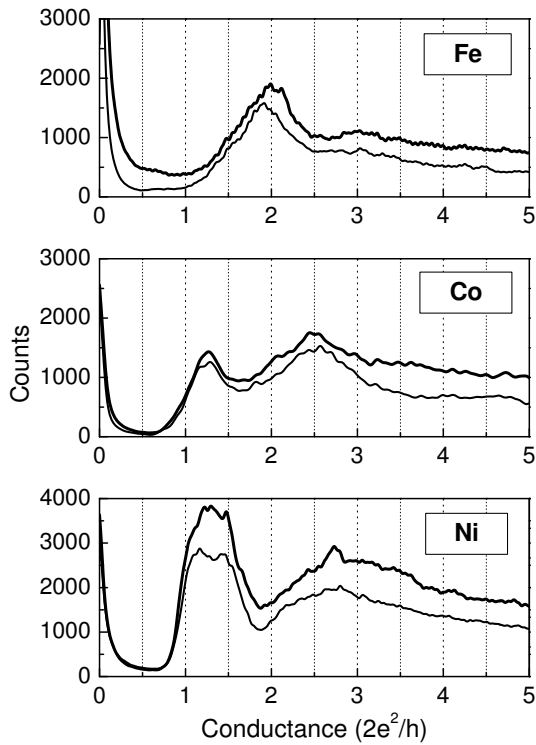} \caption
{\label{Histograms} Conductance histograms for the various metals
without magnetic field (thin curve) and when a magnetic field of
5\,T was applied (thick curve). More than a thousand traces were
used to construct each histogram and the same structure was
observed for every histogram recorded on different samples for the
same metals. The conductance was measured using a dc bias voltage
of 20\,mV.}
\end{center}
\end{figure}

As part of the experiments we have recorded traces of conductance
while pulling the contacts until breaking (seen as a sudden drop
of conductance to well-below 1\,$G_0$). Once broken the electrodes
were pushed together again until the contact was big enough so as
to assure a completely new atomic contact configuration. This is
important in order to avoid repetitive evolution in the pulling
process \cite{untiedt97}. The process of contact making followed
by controlled breaking was repeated a few thousand times and the
traces were used to build histograms of conductance such as those
shown in Fig.\,\ref{Histograms}.

We collected histograms of conductance for three magnetic metals
(Co, Fe and Ni, Fig.\ \ref{Histograms}) and a non-magnetic metal,
Pt, which has been predicted to become magnetic in a one
dimensional atomic wire configuration \cite{bahn01,delin03}. In
agreement with previous experiments done under similar conditions
\cite{sirvent96a,ludoph00a,smit01,bakker02,smit02,nielsen03}, the
histograms for the various d-metals in this work show a prominent
first peak with a value well above 1\,$G_0$ that is attributed to
the conductance of single-atom contacts. Pt is not shown here, but
looks very similar with a first peak between 1.5 and 2\,$G_0$, see
below. Below the first peak the counts rapidly drop until we find
a new rise mainly caused by the tunneling of electrons through the
vacuum barrier between the two electrodes. Note in particular that
peaks at 1\,$G_0$ and $\frac{1}{2}G_0$ are absent.

To test whether the relative orientation of magnetic domains
around the contacts plays any role in our results we have repeated
the same experiments in high magnetic fields. Prior to this we
measured the hysteresis curves for the sample wires used in the
experiments to identify the saturation field, which was for all
the cases below 2\,Tesla. In the experiments we used fields up to
5\,T, well above the saturation fields. The conductance histograms
(thin lines in Fig.\,\ref{Histograms}) show no significant
difference compared to the zero-field experiments. We even
increased the magnetic field up to 10 T in the case of Co, and
again no changes were observed.

The present results seem to contradict those obtained in the room
temperature experiments mentioned above
\cite{elhoussine02,shimizu02,gillingham02,gillingham03,gillingham03a,rodrigues03}.
Moreover, if the effects reported for room temperatures were a
result of the metals being magnetic these effects should be even
be more pronounced at low temperatures. Since this is not the case
we have looked for other explanations for the observed
(fractional) quantum peaks. The fact that all room temperature
experiments are performed under atmospheres that are considerably
less pure than that provided by cryogenic vacuum we are led to
consider the possibility of atomic-scale contamination of the
contact by foreign atoms or molecules.

For Pt it has recently been shown that a controlled contamination
of the break junction with H$_2$ leads to the appearance of a peak
in the conductance histograms situated very close to 1\,$G_0$
\cite{smit02}. It was shown that this peak in the conductance
histogram corresponds to stable configurations of a single
hydrogen molecule having a single conductance channel that is
nearly perfectly transmitting. Since a hydrogen molecule only
allows a single channel of conductance one may guess that when
such a molecule bridges two electrodes that are fully
spin-polarized the conductance would be limited to
$\frac{1}{2}G_0$. Considering that hydrogen is a rather common
contaminant it may have influenced the experiments at room
temperature and was acting as a `filter' limiting the number of
channels to one.

\begin{figure}[!t]
\begin{center}
\epsfig{width=7cm,figure=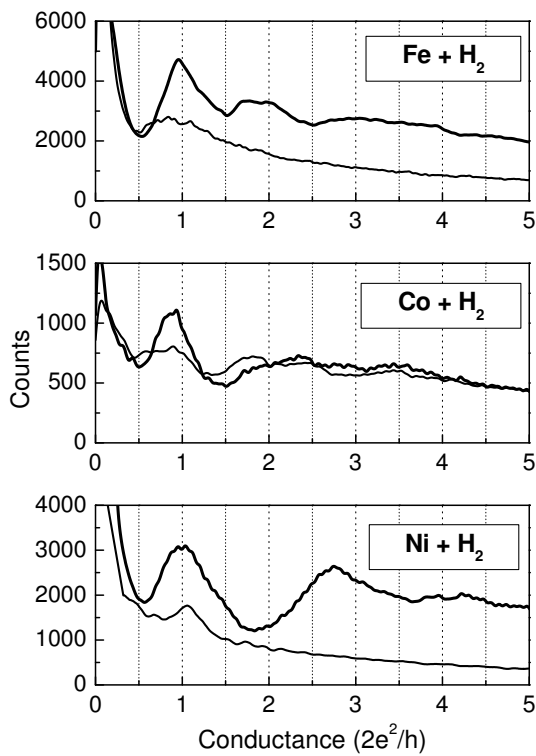} \caption
{\label{Hydrogen} Conductance histograms for the various magnetic
metals in the presence of a hydrogen atmosphere. The traces were
recorded using a bias voltage of 20\,mV (thin line) and 150\,mV
(thick line).}
\end{center}
\end{figure}

In order to test this idea we have repeated the experiments in the
presence of a hydrogen atmosphere of approximately
$10^{-6}$\,mbar. For all the experiments the sample was cooled to
4.2\,K for a day in cryogenic vacuum, at this stage we could
reproduce the histograms of Fig.\,\ref{Histograms}. Then the H$_2$
gas was introduced and the histograms changed dramatically as
shown in Fig.\,\ref{Hydrogen}.  First we note that the histograms
show counts at conductances below the first peak in an amount much
higher than for the clean metals in Fig.\,\ref{Histograms}.
Second, under similar conditions as for the experiment on Pt
\cite{smit02} the characteristic peaks for the various magnetic
metals are suppressed while a new peak appears near 1\,G$_0$ in
the conductance histogram taken at bias voltages above about
100\,mV. As argued in \cite{smit02} the role of the higher bias
voltage is to provide some local heating to evaporate away the
weakly bound excess hydrogen molecules. The results suggest that
the hydrogen-bridged configuration observed for Pt is a general
feature in these metals, but we have not yet attempted to confirm
this by phonon-spectroscopy.

A dominant conductance of 1\,$G_0$ with a hydrogen molecule
between electrodes of Co, Ni and Fe suggests that the current is
at best partially spin-polarized in these atomic-sized contacts.
As argued in Refs.\,\cite{delin03,bagrets03}, when there is a
contribution of both spin components at the Fermi energy the
conductance is still carried by two spin-channels even though the
spin subbands may be shifted considerably in energy by the
exchange interaction. The same is true for Pt. We cannot exclude
the possibility of a magnetic moment developing in atomic-sized
contacts, as has been predicted \cite{bahn01,delin03}. However,
its existence cannot be concluded exclusively based on the
appearance of peaks at fractional conductance quanta in
conductance histograms, as we will illustrate next.

We have searched for effects of other possible gas molecules that
could be present in some quantities at room temperatures or in
high vacuum. We performed these studies again at low temperatures
in a controlled atmosphere. From many experiments we have found
that it is easy to tell whether or not some contaminants have
reached the sample by just looking at the conductance histograms.
In those cases where the contaminants reached the sample there is
a large contribution to the histogram for values of conductance
below 1\,$G_0$. Above a certain quantity of contaminants the
histograms have a large smooth background that decreases with
conductance (Note the difference for the backgrounds between
Fig.\,\ref{Histograms} and the thin lines in
Fig.\,\ref{Hydrogen}). This background is reduced when increasing
the applied bias voltage and normally when it is above 200\,mV
both the new peaks and background disappear and the histograms for
the bare metals are recovered.

When introducing small amounts of carbon monoxide, CO, to Pt
atomic contacts we observed two new peaks appearing in the
conductance histogram, one near $\frac{1}{2}G_0$ and one near
1\,$G_0$ (Fig.\,\ref{Impurity CO}). The gas was introduced through
a capillary heated by sending a current through a resistive wire
running inside the full length of the capillary.

\begin{figure}[!t]
\begin{center}
\epsfig{width=7cm,figure=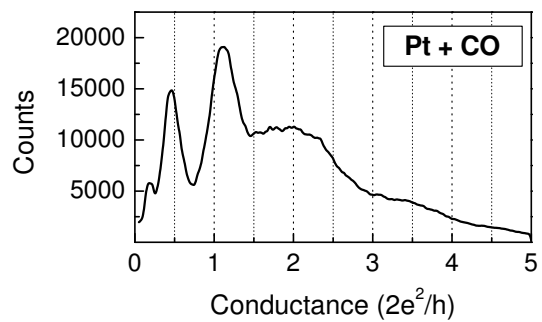} \caption {\label{Impurity
CO} Conductance histogram for a Pt atomic-sized contact
intentionally contaminated by CO. The positions of the peaks at
about the first two multiples of $\frac{1}{2}G_0$ are suggestive
of the lifting of spin degeneracy, but other explanations are
preferred. A bias voltage as high as 150\,mV was applied in order
to reduce the background.}
\end{center}
\end{figure}

These peaks behave as those from other contaminants, such as
H$_2$, and the positions are remarkably similar to those that
would be expected for half-integer quantization. There is,
however, no evidence that could relate these peaks to magnetism.
The peak near $\frac{1}{2}G_0$ likely results from a stable
molecular geometry of CO between the metal electrodes, that may
have a single spin-degenerate partially transmitted channel. We
would welcome numerical calculations to verify this. Although we
cannot claim that the reported observations of half-integer quanta
of conductance are all due to contamination of the contacts by CO,
our result shows that at least there is one kind of molecule that
would produce similar behavior. Since a direct relation of the
observed half-integer peaks to magnetism is often not provided, we
prefer to tentatively attribute the observed fractional quantum
peaks to adsorbed molecular species.

In conclusion, in contrast to previous reports on Fe, Co, Ni and
Pt we have not detected fractional conductance quantization at low
temperatures for the same metals. Our experiments show that the
conductance of pure metallic atomic-sized contacts for Fe, Co, Ni
and Pt have no significant field dependence in the
histogram-averaged conductance, and show no peaks that can be
associated with pure quantization, in agreement with recent
numerical results \cite{bagrets03,delin03}. At least part of the
previously reported fractional quantization may be explained by
the presence of foreign molecules at the surface of the studied
samples, as we have demonstrated by intentional contamination of
our samples. Our results support the idea that the magnetic state
of the samples is not related in a simple manner to its
conductance. Although the electrical current in the atomic-sized
conductors may be partially spin-polarized this is not a
sufficient condition to obtain a fractional quantum of conductance
for single-atom contacts. Only when the conductance is dominated
by a single s-channel and the exchange energy is large enough to
completely block transport through one of the spin sub-channels we
will find the sought-after half integer conductance. At the
present time, the only experiment that probably fulfills these
requirements is the one recently reported by Suderow {\it et al.}\
\cite{suderow03}, where the atomic contact is between a gold tip
and a thin gold film that is deposited on top of a half-metallic
ferromagnetic manganite.

This work is part of the research program of the ``Stichting
FOM,'' which is financially supported by NWO. C.\,U.\ acknowledges
support of the Spanish Ram{\'o}n y Cajal program of the MCyT. We
are grateful to T.\,G.\ Sorop for magnetization measurements.

\bibliographystyle{apsrev}
\bibliography{QPC_v26}

\end{document}